\documentclass[a4paper,12pt]{article}
\usepackage[pdftex]{graphicx}
\usepackage{latexsym,amsfonts,amsmath,amscd,epsfig}
\usepackage{amssymb,amsxtra,mathtools}
\usepackage{url}
\usepackage{pdflscape}
\usepackage[cp1251]{inputenc}
\usepackage{d-omega2}
\newcommand{\bec}{\begin{center}}
\newcommand{\ec}{\end{center}}
\newcommand{\bee}{\begin{equation}}
\newcommand{\ee}{\end{equation}}

\DeclareMathAlphabet{\mathbi}{OML}{cmm}{b}{it}

\begin{document}

\NEWARTICLE{english}{page-first}{page-last} 
\label{page-first}

\TITLE{K-theory and phase transitions at high energies} 

\AUTHORa{T.V. Obikhod}
\ADDRESSa{Institute for Nuclear Research, NAS of Ukraine 03680 Kiev, Ukraine}
\EMAILa{E-mail: {\sf obikhod@kinr.kiev.ua}}

\ABSTRACT{The duality between $E_8\times E_8$ heteritic string on manifold $K3\times T^2$
and Type IIA string compactified on a Calabi-Yau manifold
induces a correspondence between vector bundles on $K3\times T^2$ and
 Calabi-Yau manifolds. Vector bundles over compact
base space $K3\times T^2$ form the set of isomorphism classes, which
is a semi-ring under the operation of Whitney sum and tensor product. 
The construction of semi-ring $Vect\ X$ of isomorphism classes
of complex vector bundles over $X$ leads to the ring $KX=K(Vect\ X)$,
called Grothendieck group. As $K3$ has no isometries and
no non-trivial one-cycles, so vector bundle winding modes
arise from the $T^2$ compactification. Since we have focused
on supergravity in $d=11$, there exist solutions in $d=10$ for
which space-time is Minkowski space and extra dimensions are $K3\times T^2$.
The complete set of soliton solutions of supergravity theory is
characterized by $RR$ charges, identified by K-theory. 
Toric presentation of Calabi-Yau through Batyrev's toric
approximation enables us to connect transitions between Calabi-Yau
manifolds, classified by enhanced symmetry group, with K-theory classification.}


\HEADERS{K-theory and phase transitions at high energies}{T.V. Obikhod}

\MAKETITLE

\fancypagestyle{initpage}


\newpage
\section{Introduction}
M-theory, that first conjectured
by Edward Witten in the spring of 1995, unifies
all versions of string theory: type I, type IIA,
type IIB, and two flavors of heterotic string theory
(SO(32) and $E_8\times E_8$). Each of these five string theories
is limiting case of M-theory, and should be approximated
by eleven-dimensional supergravity at low energies. 
M-theory is connected with the AdS/CFT correspondence
and should describe branes \cite{1.}. 
For connection of M-theory with experimental data
is used the mechanism of compactification of its 
extra dimensions to four-dimensional world, 
that can be verified at the LHC.
One of the fundamental questions of theoretical 
high energy physics is the question of phase transitions
of solitonic states, like D-branes. 
Due to the higher dimensional models, ADD or RS \cite{2.}
in some scenarios involving extra dimensions of space, 
the Planck mass can be as low as the TeV range. 
As the Large Hadron Collider (LHC) has energy 14 TeV 
for proton-proton collisions it was argued \cite{3.}
that black hole production could be an important
and observable effect at the LHC or future higher-energy colliders. 
Such quantum black holes or other high energetic 
silitonic objects should decay emitting sprays 
of particles that could be seen by detectors.

	The purpose of the article is to study 
such solitonic objects with the help of effective model - D-branes. 
Our attempts for connection string theory with experiment 
are focused on the compactification of extra dimensions to $K3\times T^2$
for construction models of our four-dimensional world.  
The studied cases involve higher-dimensional spaces - D6-branes 
in for-dimensional space, for example. In section 2 we'll consider 
the equivalence of  vector bundles on $K3\times T^2$ and Calabi-Yau manifolds. 
We are dealing with a special type of Calabi-Yau manifolds, 
when it is both an elliptic bundle over Hirzebruch surface $F_{2k}$  
and K3-fibration over  the one dimensional projective space. 
In accordance with \cite{4.} there is the correspondence 
between the heterotic string compactifications on $K3\times T^2$ 
and type II string compactifications on Calabi-Yau 
threefolds of such type. This is  connected with fact 
that for compactification of the heterotic string on $K3\times T^2$ 
is used the embedding, equating the spin 
connection of the manifold with the gauge connection. 
Through examples, it is possible that the moduli spaces 
of many different N=2 heterotic vacua are connected in 
a similar way to type II string compactifications 
on Calabi-Yau. To each type of Calabi-Yau corresponds 
its enhanced symmetry according to Batyrev's toric approximation. 
In section 3 will be presented K-theory description 
of vector bundles over $K3\times T^2$. 
In section 4 we consider D-brane RR charge calculation 
corresponding to orbifold. The conclusion is connected with 
received fact that breaking of enhanced symmetries of 
Calabi-Yau to Standard model is connected with K-theory 
description of vector bundles over $K3\times T^2$, which differ 
from each other by the rank of the bundle.

\section{Calabi-Yau transitions }

From \cite{5.} we know 
that there exist duality between (0, 4) 
compactifications of the $E_8\times E_8$ heterotic 
string on the manifold $K3\times T^2$ and the type IIA 
string compactified on a Calabi-Yau manifold \cite{6.}. 
This duality corresponds to the equivalence 
of  vector bundles on $K3\times T^2$ and Calabi-Yau manifolds. 
This Calabi-Yau are of special type - K3-fibration over P$^1$ 
projective space.
In the context of M- and F-theory \cite{7.}
the dynamics on the moduli space 
of string theory would allow to determine the 
physical ground state of the string. The criterion 
which distinguishes between different vacua of 
string theory is the compactifications of these 
theories to three and four dimensions, in particular 
to CalabiYau fourfolds. Toric description of elliptic 
Calabi-Yau manifold according to Batyrev's approximation 
\cite{8.} can be realized by dual 
polyhedron  which is devided by triangle  on the top and bottom , 
For fourfolds of type 
$X_{18k+18}(1, 1, 1, 3k, 6k + 6, 9k + 9) $
the gauge groups are written in the following way \cite{9.}: 
\[H \times SU(1)\ \ \ \mbox{for} k = 1 ,\]
\[H \times SO(8)\ \ \ \mbox{for} k = 2 ,\]
\[H \times E_6 \ \ \ \mbox{for} k = 3 ,\]
\[H \times E_7 \ \ \ \mbox{for} k = 4 ,\]
\[H \times E_8 \ \ \  \mbox{for} k = 5 ,\]
\[H \times E_8 \ \ \ \mbox{for} k = 6 .\]
Thus, to each type of Calabi-Yau corresponds its enhanced symmetry.  
The moduli space of string theory vacuum 
can be deformed by vevs through breaking 
the gauge group. For example, $E_8$ can be completely broken through the chain
\[E_8 \rightarrow E_7 \rightarrow E_6 \rightarrow
SO(10) \rightarrow SU (5) \rightarrow 
SU (4) \rightarrow SU (3) \rightarrow SU (2) \rightarrow SU (1)\ .\]
So, the breaking of the gauge group is connected with 
phase transitions between different Calabi-Yau manifolds.

\section{K-theory description of vector bundles over $K3\times T2$}

As transitions between Calabi-Yau, 
described in such technique, are known, 
it would be interesting to understand an 
adequate mathematical method for description of transitions between 
different vector bundles 
over compact base space $K3\times T^2$ according to duality \cite{5.}. 
The studying of such vector bundles was performed 
by \cite{9.}, where the bundle V on $K3\times T^2$ is fixed as follows
\[V=\oplus_{i}p_{1}^{*}L_{x_{i}}\otimes p_{2}^{*}M_{i} \ ,\]
where $L_{x_{i}}$ are the line bundles on  $T^2$
corresponding to $x_{i}$ - a point in the dual 
torus $\check{T}^2$  , $p_1$ and $p_2$ are projections from 
the product $K3\times T^2$ to the factors $T^2$ and $K3$, respectively.

   As is known from \cite{11.},
vector bundles over compact base spaces form the set of 
isomorphism classes of vector bundles over $X$ (in our case $X$ is $K3\times T^2$).  
This is a semi-ring under the operations of 
Whitney sum and tensor product. It contains the disjoint union 
\[\mbox{Vect} X=\bigcup\limits_{d=0}^{\infty} \mbox{Vect}_d X\]
where Vect$_d X$ comprises the classes 
of vector bundles of rank d.  Such construction 
of semi-ring Vect $X$ of isomorphism classes of 
complex vector bundles over $X$  leads to the ring
$KX:=K(\mbox{Vect} X)$, which is called 
a Grothendieck group a contravariant functor 
from compact topological spaces to rings.
From \cite{12.}
is known that, there exists the isomorphism of 
free sheaves of rank $n$ and classes of 
vector bundles of rank $n$ .  
The category Vec$_r(X)$ of vector bundles 
of rank r on $X$ and the category Loc$_r(X)$ of 
locally free sheaves of rank r on $X$ are 
equivalent, as defined by Hartshorne, \cite{11.}. 
In \cite{13.} is written about D-brane as 
locally free sheaf. From \cite{14.} it has 
been observed (for example, as in \cite{15.}) 
that branes supported on complex submanifolds 
of complex varieties are naturally described 
in terms of coherent sheaves. Therefore, Grothendieck 
groups of coherent sheaves, the holomorphic 
version of K-theory, can be used to describe 
D-branes in the case that all D-branes are 
wrapped on complex submanifolds.

\section{String theory and RR charge}

Since we are dealing with M-theory theory 
of five string theories, it is necessary to 
say that string looks like an ordinary particle, 
with its mass and charge. A physical object 
that generalizes the notion of a point particle 
to higher dimensions is a brane. A point particle 
can be viewed as a brane of dimension zero, while 
a string can be viewed as a brane of dimension one. 
It is also possible to consider higher-dimensional 
branes. In dimension p, they are called p-branes. 
Branes are dynamical objects which can propagate 
through spacetime according to the rules of quantum 
mechanics. They can have mass and other attributes 
such as RR-charge. Much of the current research in 
M-theory attempts to better understand the properties 
of these branes. So, the studying of D-brane 
classification with the help of K-theory description 
of RR-charge is of great importance.

  Generally believed of K-theory as a ``poor man's derived category''
that knows only about D-brane charge \cite{13.}. 
A D-brane charge corresponding to a vector bundle E is given by formula
\[Q(E)=ch(E)\sqrt{td(X)}\ ,\]
where $ch(E)$ is the Chern character of $E$ 
and $td(X)$ is the Todd class of the tangent 
bundle of $X$. The charge of a D-brane given 
by a coherent sheaf can be computed using the
Grothendieck-Riemann-Roch theorem \cite{13.}.

  As is stressed in \cite{16.} 
blowing-up of $T^4/Z_2$ is K3 space. Therefore, instead 
of $K3\times T^2$ space we can consider $T^6/Z_2$ space - orbifold. 
D-brane that passes through an orbifold fixed 
point carries RR charge. In a supergravity 
approximation we can take the large volume limit to
describe the backgrounds and these are RR  backgrounds 
\cite{17.}. 
Large volume charges  are connected with RR charges by formula:
\[Q_4=n_1-2n_2+n_3,\ Q_2=-n_1+n_2,\ Q_0=\frac{n_1+n_2}{2}\]
which define the Chern character 
\[ch(n_1n_2n_3)=Q_4+Q_2\omega+Q_0\omega^2\ ,\]
$\omega$ is Wu parameter.
The rank of the vector bundle $E$ is $Q_4$, $Q_2=c_1,\ Q_0=\frac{c_1^2}{2}$, where 
$c_1$ is the first Chern class.
From \cite{10.} the first Chern class of $SU(N)$ bundle
$E$ over $K3\times T^2$, for example, is zero, therefore $Q_2=0$ and $Q_0 =0$ 
and vector bundles over $K3\times T^2$ are differ through the rank of the bundle. 

\section{Conclusion}
The conclusion is connected with the fact
that morphisms of distinct Calabi-Yau permit
an interpretation in terms of topological K theory 
or Grothendieck groups. In spite of the fact
that $K3~T^4/Z_2$ and  the full space $K3\times T^2~T^6/Z_2$,
we can't use the notion of Aspinwall \cite{13.} 
that D-branes on the orbifold $C_d/G$ and open strings 
between them are described
by the derived category of McKay quiver 
representations (with relations) because 
we cannot have the derived category of a 
``compact'' CY manifold represented by a quiver.
Anyway, in spite of the fact that K-theory contains 
much less information than the derived category this 
is the beginning for understanding of  brane 
classification when the symmetry is broken from E$_8$ 
to the Standard molel. According to \cite{18.} 
when we focus our attention on supergravity in d=11, 
solutions also exist for N=1 in d=10, 9 and 8 dimensions 
for which spacetime is Minkowski space and for which 
the extra dimensions are $K3\times T2$, 
$K3\times S^1$ and $K3$, respectively.  
Starting from 64+64 components of N=1 in d=10, 
we obtain 96+96 components in d=4 of N=2 
supergravity and so on to the Standard model.


\noindent\textbf{T.V. Obikhod}\\

\end{document}